\documentclass[iop]{emulateapj}

\usepackage{natbib}
\shorttitle{Thirteen New \ion{He}{2} Quasars}
\shortauthors{Syphers et al.}

\begin{document}

\title{{\em HST}/COS\altaffilmark{1} Observations of Thirteen New \ion{He}{2} Quasars}

\author{David Syphers\altaffilmark{2},
Scott F.\ Anderson\altaffilmark{3},
Wei Zheng\altaffilmark{4},
Avery Meiksin\altaffilmark{6},
Donald P.\ Schneider\altaffilmark{7,8},
Donald G.\ York\altaffilmark{9,10}
}

\altaffiltext{1}{Based on observations with the NASA/ESA \textit{Hubble Space Telescope} obtained at the Space Telescope Science Institute, which is operated by the Association of Universities for Research in Astronomy, Inc., under NASA contract NAS5-26555.}

\altaffiltext{2}{CASA, Department of Astrophysical and Planetary Sciences, University of Colorado, Boulder, CO 80309, USA; David.Syphers@colorado.edu}

\altaffiltext{3}{Department of Astronomy, University of Washington, Seattle, WA 98195, USA}

\altaffiltext{4}{Department of Physics and Astronomy, Johns Hopkins University, Baltimore, MD 21218, USA}

\altaffiltext{6}{Scottish Universities Physics Alliance (SUPA), Institute for Astronomy, University of Edinburgh, Royal Observatory, Edinburgh EH9 3HJ, UK}

\altaffiltext{7}{Department of Physics \& Astronomy, Pennsylvania State University, University Park, PA 16802, USA}

\altaffiltext{8}{Institute for Gravitation and the Cosmos, Pennsylvania State University, University Park, PA 16802, USA}

\altaffiltext{9}{Department of Astronomy and Astrophysics, The University of Chicago, Chicago, IL 60637, USA}

\altaffiltext{10}{Enrico Fermi Institute, The University of Chicago, Chicago, IL 60637, USA}

\begin{abstract}
The full reionization of intergalactic helium was a major event in the history of the IGM, and UV observations of the \ion{He}{2} Gunn-Peterson trough allow us to characterize the end of this process at $z \sim 3$.
Due to intervening hydrogen absorption, quasars allowing such study are rare, with only 33 known in the literature, and most of those are very recent discoveries.
We expand on our previous discovery work, and present 13 new \ion{He}{2} quasars with redshifts $2.82 < z < 3.77$, here selected with $\sim$80\% efficiency, and including several that are much brighter than the vast majority of those previously known.
This is the largest sample of uniformly observed \ion{He}{2} quasars covering such a broad redshift range, and they show evidence of IGM opacity increasing with redshift, as expected for the helium reionization epoch. 
No evidence of \ion{He}{2}~Ly$\alpha$ quasar emission is seen in individual or averaged spectra, posing a problem for standard models of the broad line region.
The current rapid advance in the study of \ion{He}{2} quasars has been greatly facilitated by the Cosmic Origins Spectrograph on {\it HST}, and we discuss the instrumental and other subtleties that must be taken into account in IGM \ion{He}{2} observations.
\end{abstract}


\section{Introduction}
\label{sec:intro}

The epoch of full helium reionization was a major transition for the intergalactic medium (IGM).
It likely began at $z>3.5$, possibly as early as $z \gtrsim 4$ \citep{lidz10,becker11,bolton12}, and depending on the sightline it completed at $z \sim 2.7$--$3.3$, as determined by the transition from a \ion{He}{2} Gunn-Peterson trough \citep{gunn65} to a \ion{He}{2}~Ly$\alpha$ forest \citep[e.g.,][]{shull10,syphers11a,worseck11a}.
Reionization of helium was thus delayed versus that of hydrogen at $z>6$ \citep{fan06}, as it required higher-energy photons produced by quasars, which did not become numerous until later \citep{hopkins07}.
The most useful and direct probe of the helium reionization epoch has proved to be observations of the \ion{He}{2}~Ly$\alpha$ Gunn-Peterson trough, along very rare sightlines that are free of substantial hydrogen absorption and thus show quasar continua extending to the \ion{He}{2}~Ly$\alpha$ break ($304$~\AA\ rest frame).
Only a few percent of $z \sim 3$ quasar sightlines have small enough \ion{H}{1} column densities to allow such observations \citep{moller90,zheng05}, and finding these ``\ion{He}{2} quasars'' has been a laborious task.

Other methods of observing the epoch of \ion{He}{2} reionization are more indirect, examining either the \ion{H}{1} Ly$\alpha$ forest or metal systems.
Helium reionization injects a substantial amount of energy into the IGM, which may be observed as broader \ion{H}{1} Ly$\alpha$ forest lines \citep[e.g.,][]{ricotti00,schaye00,becker11} or lower average Ly$\alpha$ forest opacity \citep[e.g.,][]{bernardi03,faucher-giguere08}.
Variations of forest line width with redshift are disputed \citep{mcdonald01,kim02,meiksin10}, and while there appears to be a real opacity dip at $z \simeq 3.2$, its interpretation is not straightforward \citep{bolton09a,mcquinn09}.
Measurements of IGM metal line widths could break the thermal/non-thermal broadening degeneracy and yield firmer indirect constraints, but these are as yet observationally impossible in the necessary density regime \citep{meiksin10}.
Optical depth ratios of metal species whose ionization potentials straddle the \ion{He}{2} Lyman limit (54.4~eV, 228~\AA) may show a change in the UV background in $z \sim 3$--4 that is possibly associated with helium reionization \citep{songaila98,agafonova07}, but this observational result is controversial \citep{kim02a,aguirre04}, and such a change is not predicted in some models due to inhomogeneities in the UV background or in metallicity \citep{furlanetto09a,bolton11}.
Helium Gunn-Peterson trough studies, while unable to probe the early stages of helium reionization \citep[$x_{\rm He\,II} \gtrsim 10$\%;][]{mcquinn09a,syphers11b}, are therefore the most direct probe of its later stages.

For many years, only a few \ion{He}{2} quasars were known, but techniques cross matching large quasar catalogs with {\it GALEX} UV photometry have greatly increased the efficiency of finding such quasars, from $\sim$3--5\% to $>$50\% \citep{syphers09b,syphers09a,worseck11}, and now in this paper to $\sim$80\%.
Verification spectra of quasars on these cross-match lists requires observed-frame far-UV (FUV) observations, and for many years prior to the installation of the Cosmic Origins Spectrograph \citep[COS;][]{green11,osterman11}, far-UV observations were limited to the Advanced Camera for Surveys/Solar Blind Channel (ACS/SBC) prism and $\lambda > 1250$~\AA.
This meant that \ion{He}{2} quasars could be observed only at redshifts $z > 3.1$, and with very low spectral resolution ($R \sim 50$--$400$).
With some notable exceptions, many of the \ion{He}{2} quasars verified with the ACS prism are faint, and thus less attractive targets for long COS observations.
This is particularly true because COS must acquire targets in the near UV, where these quasars tend to be fainter, and where COS has relatively high backgrounds.
(However, these targets are not nearly so faint in the FUV as initially thought.
The ACS flux calibration was probably a factor of $\sim$3 low; see Section \ref{sec:obs}.)

The advent of COS has allowed substantial progress, from very high quality reobservations of known \ion{He}{2} quasars \citep{shull10} to the discovery of some new lower-redshift \ion{He}{2} quasars \citep[$z \lesssim 3.0$;][]{worseck11a}.
However, the number of \ion{He}{2} quasars known remains small, and since many of those are fairly faint or at low redshift (and thus have very little of their Gunn-Peterson troughs and helium Ly$\alpha$ forests covered by COS), it is highly desirable to find new \ion{He}{2} quasars.
There are strong variations between sightlines, seen in the well-known \ion{He}{2} quasars HE2347$-$4243 \citep{reimers97,shull10}, HS1700$+$6416 \citep{davidsen96,fechner06}, Q0302$-$003 \citep{jakobsen94,heap00}, and recently reconfirmed \citep{worseck11a}, so substantial numbers are necessary to draw general conclusions.

In this paper we report the discovery of 13 new \ion{He}{2} quasars, with redshifts ranging from $2.82$ to $3.77$, including targets much brighter in the far UV than many of the previously known \ion{He}{2} quasars.
A number of these are well suited to more detailed investigation with longer COS observations.
In Section \ref{sec:obs} we discuss the observations, and instrumental subtleties including geocoronal emission and background subtraction (the appendix gives more detail on the latter point).
Section \ref{sec:results} presents the results of our search for new \ion{He}{2} quasars, and the significance of our success using our UV--optical cross-correlation technique to target such quasars.
We conclude in Section \ref{sec:conclusion}.

\section{Observations}
\label{sec:obs}

We selected targets from the catalogs of \citet{syphers09b}, which were compiled from cross-correlation of optical quasar catalogs (primarily the Sloan Digital Sky Survey [SDSS]; \citealp{york00}) with {\it GALEX} UV catalogs \citep{morrissey07}.
Attempts at finding \ion{He}{2} quasars relying only on optical catalogs have success rates of only $\sim$5\% \citep[e.g.,][]{zheng05}, while UV--optical cross-correlation catalogs boost that to $\sim$50\% \citep{syphers09b}, or, using the more refined criteria of this work, to $\sim$80\%.
We briefly describe our selection criteria here; for more details, see \citep{syphers09b,syphers09a}.

We begin with catalogs of spectroscopically confirmed quasars, using the SDSS DR7 quasar catalog \citep{schneider10} and the heterogenous compilation of \citet{veron-cetty10}.
These quasars are then matched to the {\it GALEX} GR4$+$5 UV source catalog, using a maximum match radius of 3$''$.
We match separately to FUV and NUV sources---the FUV band covers $1340$~\AA~$ \lesssim \lambda \lesssim 1790$~\AA, with an effective wavelength of $1539$~\AA, and the NUV band covers $1770$~\AA~$\lesssim \lambda \lesssim 2830$~\AA, with an effective wavelength of $2316$~\AA\ \citep{morrissey07}.
Matching {\it GALEX} to Monte Carlo quasar catalogs that have arcminute-scale offsets from their true positions, we estimated a background false-match rate of $\simeq$30\% in our 3$''$ catalog.
In practice, we preferred targets at $r<2''$ to reduce our false matches, but $\sim$10\% of true matches are located in $2'' < r < 3''$, and given the rare nature of \ion{He}{2} quasars, we choose some higher-quality candidates in this region.
Beyond $3''$, there are a negligible number of true matches.
In previous work we required detection only in one UV band, which means that a few matches were spurious UV sources, while others were true NUV matches that lacked an FUV counterpart due to an intervening hydrogen Lyman-limit system (LLS).
For the survey in this work, we required matches to both {\it GALEX} UV bands, boosting our success rate to $\sim$80\% (discussed in more detail in Section \ref{sec:match_rate}).

Note that the {\it GALEX} imaging data consist of three primary surveys of varying depth.
The shallow all-sky imaging survey (AIS) covers a large area, but some of our fainter \ion{He}{2} quasars are below its detection limit ($\sim$6~$\times 10^{-17}$~erg~s$^{-1}$~cm$^{-2}$~\AA$^{-1}$ at 3$\sigma$, and for 5$\sigma$, $\sim$50~$\times 10^{-17}$~erg~s$^{-1}$~cm$^{-2}$~\AA$^{-1}$).
Some \ion{He}{2} quasars may therefore go undetected if they are covered only by AIS imaging, and not by the medium or deep imaging surveys (MIS and DIS).
This is not a concern for low redshift ($2.7 \lesssim z \lesssim 3$), where such faint \ion{He}{2} quasars would not add much to our knowledge, but it is a concern for high redshift ($z \gtrsim 3.5$), where even our best confirmed \ion{He}{2} quasars are relatively faint.

\begin{figure*}
\epsscale{1.0}
\plotone{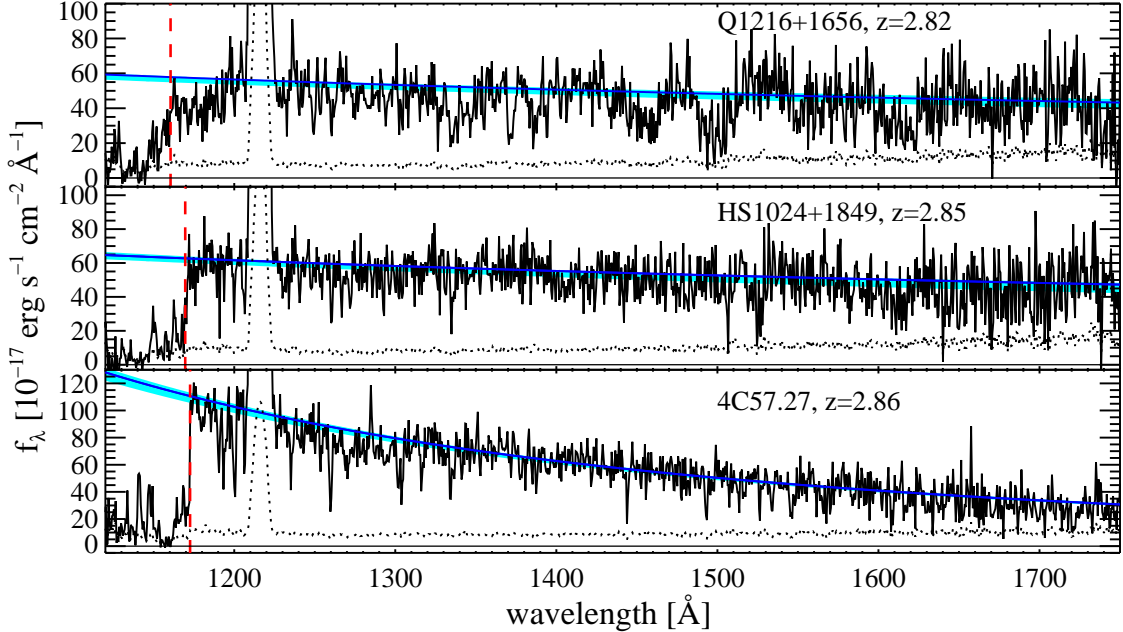}
\caption{{\it HST}/COS G140L spectra of our reconnaissance targets. The red dashed vertical line indicates the \ion{He}{2}~Ly$\alpha$ break for the redshift of the quasar. The thin horizontal solid line marks zero flux, and the dotted curve is the error spectrum. The best-fit continuum is shown in blue, and the range in possible continuum fits is shaded in cyan. Where available, night-only data has been used in regions contaminated by geocoronal line emission.}
\label{fig:spec1}
\end{figure*}

\begin{figure*}
\epsscale{1.0}
\plotone{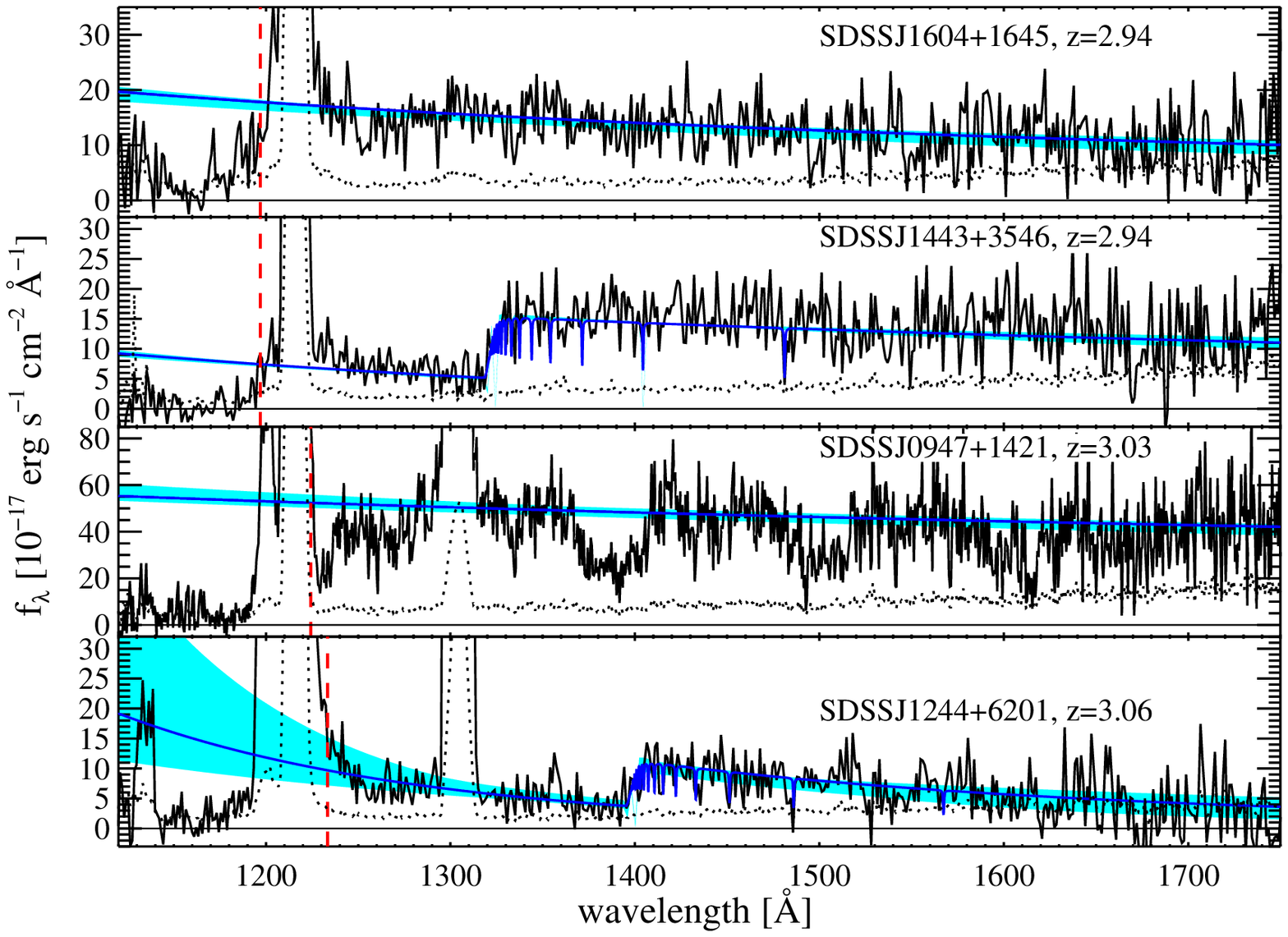}
\caption{{\it HST}/COS G140L spectra of our reconnaissance targets. The red dashed vertical line indicates the \ion{He}{2}~Ly$\alpha$ break for the redshift of the quasar. The thin horizontal solid line marks zero flux, and the dotted curve is the error spectrum. The best-fit continuum is shown in blue, and the range in possible continuum fits is shaded in cyan. Where available, night-only data has been used in regions contaminated by geocoronal line emission.}
\label{fig:spec2}
\end{figure*}

\begin{figure*}
\epsscale{1.0}
\plotone{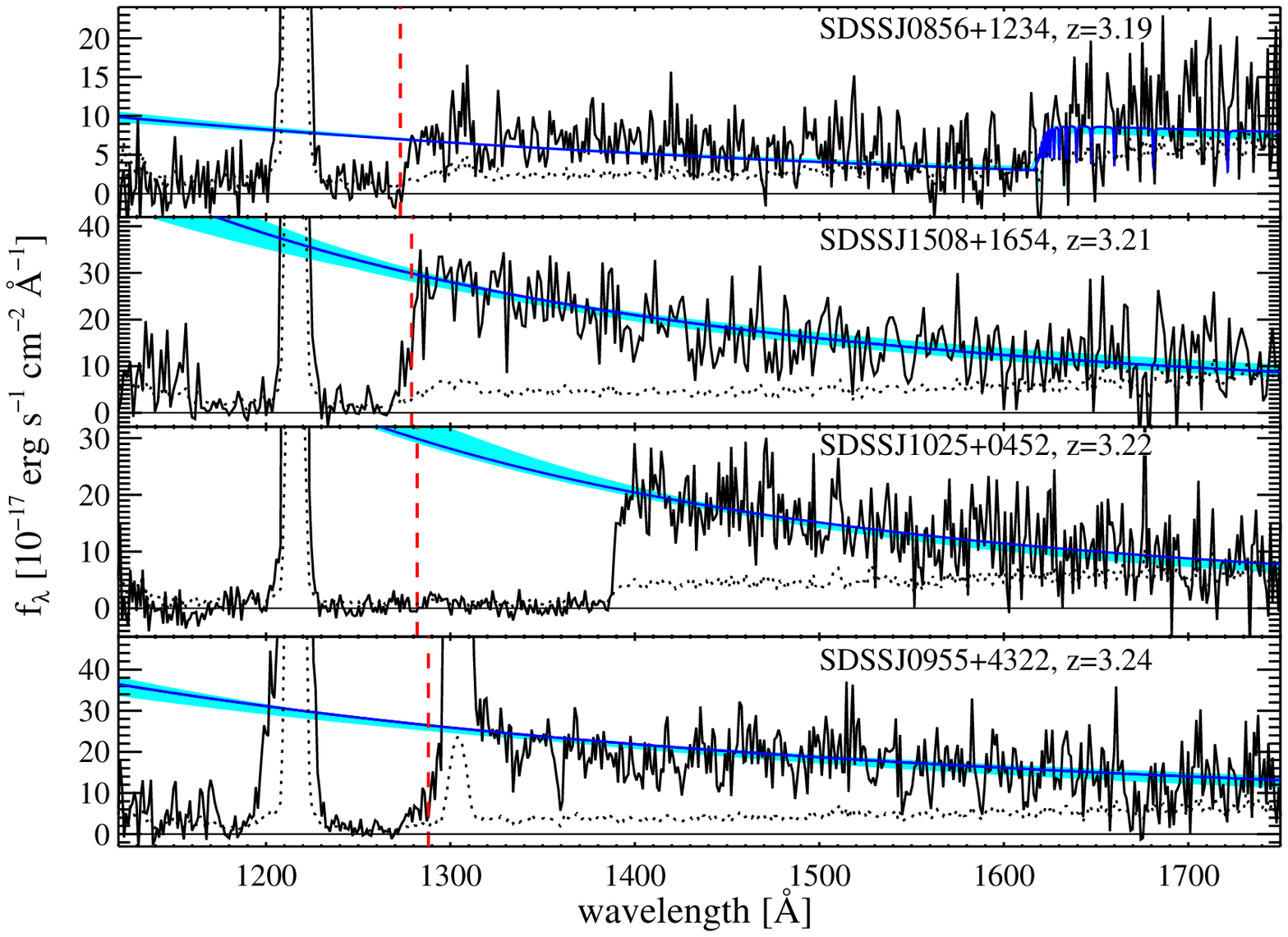}
\caption{{\it HST}/COS G140L spectra of our reconnaissance targets. The red dashed vertical line indicates the \ion{He}{2}~Ly$\alpha$ break for the redshift of the quasar. The thin horizontal solid line marks zero flux, and the dotted curve is the error spectrum. The best-fit continuum is shown in blue, and the range in possible continuum fits is shaded in cyan. Where available, night-only data has been used in regions contaminated by geocoronal line emission.}
\label{fig:spec3}
\end{figure*}

\begin{figure*}
\epsscale{1.0}
\plotone{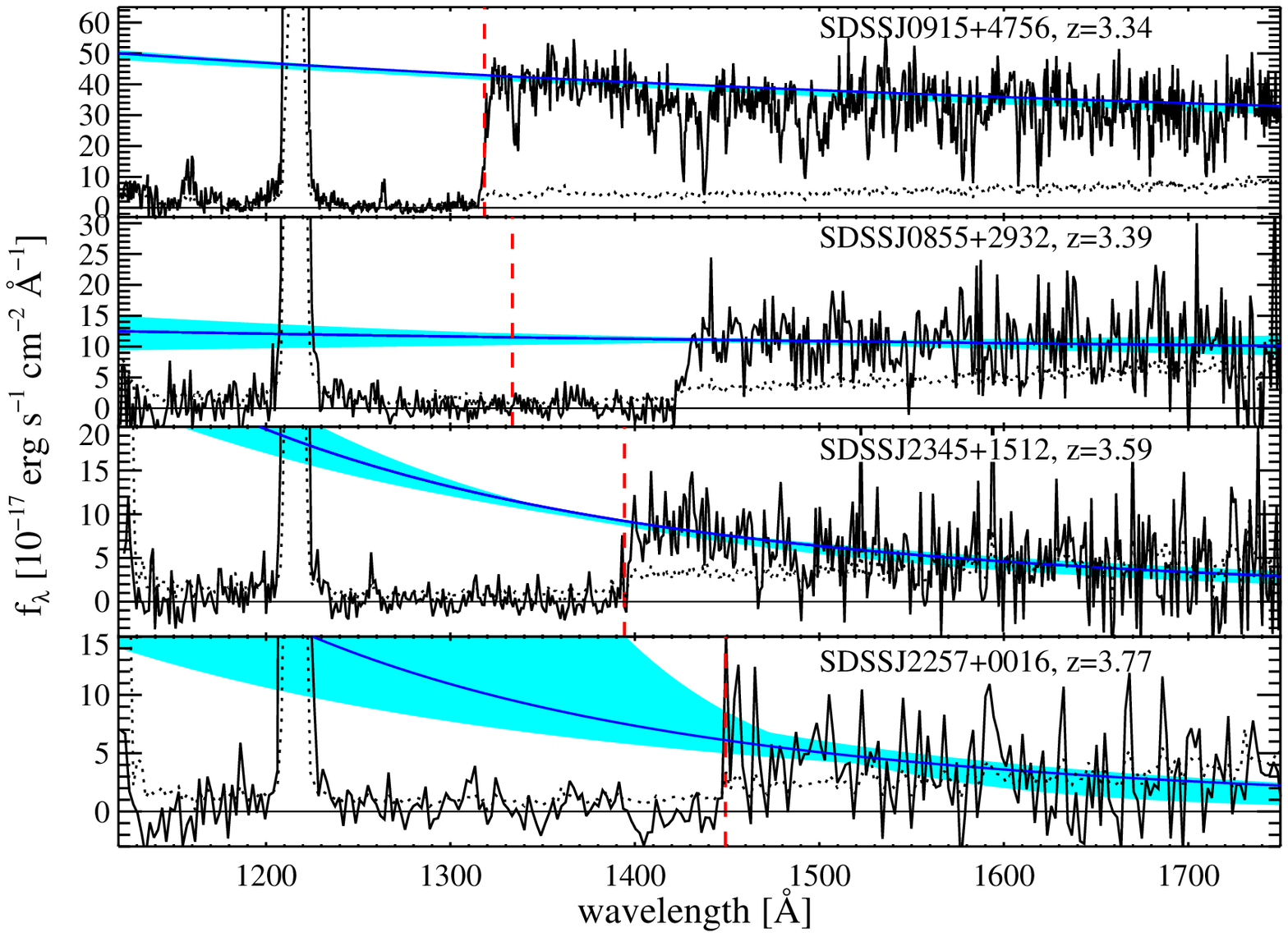}
\caption{{\it HST}/COS G140L spectra of our reconnaissance targets. The red dashed vertical line indicates the \ion{He}{2}~Ly$\alpha$ break for the redshift of the quasar. The thin horizontal solid line marks zero flux, and the dotted curve is the error spectrum. The best-fit continuum is shown in blue, and the range in possible continuum fits is shaded in cyan. Where available, night-only data has been used in regions contaminated by geocoronal line emission.}
\label{fig:spec4}
\end{figure*}

In {\it HST} GO 12178, we observed 16 potential \ion{He}{2} quasars from our list of candidates with the low-resolution FUV grating of COS, G140L ($R \sim 2000$--3000), most for one orbit and some for two.
The data were reduced using CALCOS 2.13.6.
We coadded exposures with custom software that flatfields the data, taking into account spectrum defects like the dips due to shadows from the grid of wires placed above the detector to improve its quantum efficiency.
Observation details are presented in Table~\ref{tab:obs}, and the spectra are shown in Figures~\ref{fig:spec1}--\ref{fig:spec4}.

\begin{deluxetable*}{lllcrrclc}
\tablecolumns{9}
\tablewidth{0pc}
\tabletypesize{\footnotesize}
\tablecaption{{\it HST}/COS observations}
\tablehead{
\colhead{Name} & \colhead{R.A.} & \colhead{Decl.} & \colhead{Redshift} & \colhead{FUV $f_{\lambda}$} & \colhead{$\alpha_{\nu}$\tablenotemark{b}} & \colhead{Exp. Time} & \colhead{Obs. Date} & \colhead{$\tau_{{\rm eff, \, He \, II \; Ly}\alpha}$} \\
\colhead{} & \colhead{(J2000)} & \colhead{(J2000)} & \colhead{} & \colhead{($10^{-17}$)\tablenotemark{a}} & \colhead{} & \colhead{(s)} & \colhead{} & \colhead{}}
\startdata
LBQS 1216+1656 & 184.83501 & 16.65820 & 2.82 & $57.9^{+0.5}_{-3.0}$ & $-1.29^{+0.11}_{-0.10}$ & 2008 & 2011 May 25 & $1.83^{+0.12}_{-0.15}$ \\
HS 1024+1849 & 156.89222 & 18.57433 & 2.85 & $62.7^{+1.2}_{-2.7}$ & $-1.30^{+0.20}_{-0.09}$ & 2038 & 2011 Mar 14 & $2.01^{+0.17}_{-0.15}$ \\
4C57.27 & 240.98303 & 57.51511 & 2.85 & $110.5^{+2.7}_{-4.9}$ & $1.20^{+0.06}_{-0.06}$\tablenotemark{c} & 2408 & 2011 Mar 4 & $1.95^{+0.08}_{-0.10}$ \\
SDSS J160441.47+164538.3 & 241.17280 & 16.76064 & 2.94 & $17.8^{+0.4}_{-1.5}$ & $-0.49^{+0.41}_{-0.26}$ & 2066 & 2011 Apr 30 & $1.94^{+0.27}_{-0.29}$ \\
SDSS J144311.58+354646.3 & 220.79829 & 35.77954 & 2.94 & $7.4^{+0.2}_{-0.3}$ & $-0.82^{+0.53}_{-0.34}$ & 1990 & 2011 Jun 9 & $2.75^{+1.03}_{-0.55}$ \\
SDSS J094734.19+142116.9 & 146.89249 & 14.35471 & 3.03 & $52.4^{+3.6}_{-2.1}$ & $-1.39^{+0.29}_{-0.01}$ & 2116 & 2011 Apr 27 & $2.60^{+0.26}_{-0.19}$ \\
SDSS J124456.98+620143.0 & 191.23743 & 62.02862 & 3.06 & $9.7^{+5.5}_{-2.5}$ & $3.14^{+3.99}_{-2.14}$ & 1906 & 2011 May 6 & $2.18^{+1.01}_{-0.67}$ \\
SDSS J085633.57+123428.5 & 134.13989 & 12.57459 & 3.19 & $7.0^{+0.2}_{-0.1}$ & $-0.63^{+0.37}_{-0.28}$ & 2012 & 2010 Oct 8 & $1.92^{+0.38}_{-0.28}$ \\
SDSS J150828.78+165433.1 & 227.11992 & 16.90920 & 3.21 & $29.9^{+1.1}_{-1.7}$ & $1.93^{+0.54}_{-0.67}$ & 996 & 2011 Sep 9 & $3.09^{+0.43}_{-0.35}$ \\
SDSS J102509.63+045246.7 & 156.29016 & \phn 4.87966 & 3.22 & $29.9^{+4.3}_{-0.8}$ & $2.33^{+0.98}_{-0.06}$ & 1500 & 2011 Mar 27 & \nodata \\
SDSS J095546.35+432244.7 & 148.94316 & 43.37909 & 3.24 & $26.5^{+0.1}_{-1.4}$ & $0.29^{+0.45}_{-0.27}$ & 1846 & 2010 Oct 7 & $3.40^{+0.38}_{-0.34}$ \\
SDSS J091510.01+475658.7 & 138.79172 & 47.94966 & 3.34 & $43.0^{+0.2}_{-1.5}$ & $-1.07^{+0.21}_{-0.10}$ & 5520 & 2010 Oct 6 & $>6.84$ \\
SDSS J085503.81+293248.9 & 133.76588 & 29.54693 & 3.39 & $11.5^{+0.6}_{-1.2}$ & $-1.53^{+0.78}_{-0.98}$ & 1728 & 2010 Nov 28 & \nodata \\
SDSS J005401.48+002847.7 & \phn 13.50618 & \phn 0.47994 & 3.41 & \nodata\phn\tablenotemark{d} & \nodata\phn\phn & 1796 & 2010 Nov 3 & \nodata \\
SDSS J234522.19+151217.3 & 356.34248 & 15.20482 & 3.59 & $9.2^{+0.1}_{-0.6}$ & $3.06^{+1.47}_{-0.88}$ & 1425\tablenotemark{e} & 2010 Oct 13 & $5.04^{+\infty}_{-1.87}$ \\
SDSS J225759.67+001645.6 & 344.498625 & \phn 0.279342 & 3.77 & $6.1^{+2.5}_{-1.1}$ & $3.35^{+9.26}_{-1.34}$ & 1415\tablenotemark{e} & 2010 Oct 16 & $>3.75$ \\
\enddata
\tablenotetext{a}{erg~s$^{-1}$~cm$^{-2}$~\AA$^{-1}$, from the fit to the COS spectrum at the \ion{He}{2} break, $(1+z_{\rm em}) \times 304$~\AA, or LLS break, $(1+z_{\rm LLS}) \times 912$~\AA, if not a \ion{He}{2} quasar.}
\tablenotetext{b}{Defined $f_{\nu} \propto \nu^{\alpha_{\nu}}$. This is the spectral index redward of any LLS seen in the COS data, but does not take into account LLS in the near UV ($1800 \lesssim \lambda \lesssim 3800$~\AA). See the text for the derivation of the error estimates.}
\tablenotetext{c}{The $z=0.96$ LLS is {\it not} taken into account here; above the break, the slope is not well constrained, but is probably $\alpha_{\nu} \sim -0.5$.}
\tablenotetext{d}{This quasar is not the source of the FUV light.}
\tablenotetext{e}{We report the actual science exposure time; this does not include failed exposures from guide star acquisition problems.}
\label{tab:obs}
\end{deluxetable*}

We adopt 7 pixels ($0.56$~\AA) as a resolution element, since this is approximately the FWHM of the line-spread function (LSF).
Note that the on-orbit LSF is non-Gaussian, due primarily to mirror polishing errors on {\it HST} \citep{kriss11}.
For our higher-S/N cases, we plot the spectra binned to the standard 7~pixels, but in lower-S/N cases we use 14~pixels, and for our lowest-flux target (SDSSJ2257+0016) we use 28~pixels.
The error spectra are calculated from the CALCOS pipeline error estimates assuming that the noise $\sigma \propto N^{-0.5}$, where $N$ is the bin size in pixels, which would be the case for truly independent noise in each pixel.
COS has fixed-pattern noise features that make this assumption of independence an imperfect one.
Medium-resolution COS spectra with moderate to high S/N show that binned noise may scale as $N^{\alpha}$ with $\alpha \simeq -0.37$ (Keeney et al.\ 2012, in prep.), although our own direct tests of our low-S/N G140L data show that $\alpha \lesssim -0.44$, closer to the assumption of independent noise.
This binning uncertainty means we may be underestimating the noise in our spectra by $\lesssim$10\%, but this error has little impact on our results.
We use the error spectra only for fitting the continua---in particular, we do not use it when calculating Gunn-Peterson optical depths.

We note in passing that the flux calibration of the ACS/SBC PR130L prism is likely incorrect by a significant amount.
Zheng et al.\ (2012, in prep.) observe four quasars with COS that had prior high-S/N observations with the ACS/SBC prism \citep{zheng08,syphers09b,syphers09a}.
It is notable that in every case, the flux observed with COS is approximately three times higher than that observed with ACS.
While quasars are variable, particularly at shorter wavelengths, it would be quite unlikely for all these quasars to vary in the same direction and in about the same proportion.
Low-quality spectroscopic observations of two of these objects with STIS \citep{zheng05} and imaging observations of all four with {\it GALEX} also suggest fluxes in agreement with COS rather than ACS.
It therefore seems likely that the flux calibration of the ACS/SBC prisms \citep{larsen06} is unreliable at the faint end.

\subsection{Removing Geocoronal Emission}

Geocoronal emission-line contamination, primarily \ion{H}{1}~Ly$\alpha$ and \ion{O}{1}~$\lambda$1302, can be quite substantial in the far UV.
This contamination is a problem for identifying quasar emission lines and assessing the nature of the Gunn-Peterson absorption in higher-redshift quasars.
Our data from COS is collected in TIME-TAG mode, where every count has a time stamp, and we can therefore extract the spectra during orbital night, when geocoronal emission is substantially weaker.
In practice, we cut on whether or not \ion{O}{1}~$\lambda$1302 is visible above the background---this not only tracks the day/night divide well, and implicitly accounts for any increased background near the South Atlantic Anomaly (SAA), but \ion{O}{1} emission is actually the quantity we are directly interested in.
The spectra shown in Figures~\ref{fig:spec1}--\ref{fig:spec4} use all available data except in those wavelength regions contaminated by geocoronal emission, where night-only data is used.
We verify below that geocoronal line photons scattered to other wavelengths are insignificant even in the Gunn-Peterson trough, and thus this analysis technique is justified.

The strongest geocoronal line is \ion{H}{1}~Ly$\alpha$, which is always quite strong, although an order of magnitude weaker at night.
\ion{O}{1}~$\lambda$1302 is prominent during the day, but almost entirely absent at night.
Nonetheless, we avoid Gunn-Peterson measurements in the region contaminated by this line, as there is some small residual flux.
Weaker and more rarely seen features are geocoronal \ion{N}{1}~$\lambda$1134 and \ion{O}{1}~$\lambda$1356, but they are common in continuous-viewing-zone (CVZ) observations.
These lines are relatively weak, and entirely negligible during night-only periods.
\ion{N}{1}~$\lambda$1200 is strong during high-geocoronal periods, but in G140L this line is blended with \ion{H}{1}~Ly$\alpha$, and we always avoid the entire region.
We may see, very weakly, \ion{N}{1}~$\lambda$1493, but this is not in the Gunn-Peterson trough for any object, and thus has negligible impact on our work.
We do not detect other geocoronal lines like \ion{O}{1}~$\lambda$1152 \citep[mentioned in][]{dixon10}.

Outside of regions contaminated by geocoronal emission, dark current from the detector is the dominant source of background ($\sim$$10^{-4}$~counts~s$^{-1}$~pix$^{-1}$ in the one-dimensional spectrum with the default extraction window).
Earthshine, Zodiacal light, and scattered light from geocoronal lines are all small sources of background, compared to the dark current.
We test the effect of scattered geocoronal emission by calculating the optical depths in very black troughs using all data for an object, and then using only that data collected during orbital night.
We see no effect other than larger error bars at night from the smaller quantity of data.
Earthshine and zodiacal light are at least five orders of magnitude fainter than dark current at all FUV wavelengths \citep{dixon10}.

\subsection{Verifying Background Subtraction}
\label{sec:bg}

When analyzing very dark troughs, where the source is comparable to or fainter than the background, subtraction of that background is a serious issue.
Two approaches have been taken the literature regarding COS background: \citet{worseck11a} have assumed a constant background, while \citet{syphers11b} have used the pixel-by-pixel background CALCOS derives for each exposure from background windows offset in the cross-dispersion direction.
The first approach is suboptimal because the background is noticeably variable both in time and in position (in both dispersion and cross-dispersion directions)---see Appendix~\ref{app:bg}.
While the second method attempts to take this variability into account, the very low dark rate makes this problematic, as detailed below.

There are a number of potential concerns with the COS background estimate: 
(1) the background is estimated from pixels that are unexposed, and thus only measures dark current, not earthshine, scattered airglow lines, or Zodiacal light;
(2) the background windows in their default locations include some actual airglow photons;
(3) the background is lower in the spectrum extraction region than in the region used to estimate the background, which is offset in the cross-dispersion ($y$) direction (thus this feature in the background is called ``$y$-dip'');
(4) the background varies as a function of $x$, the dispersion direction;
(5) the background is time dependent, both on short scales due to the local orbital environment, and on longer scales due to changes in sensitivity or instrumental settings;
(6) there are simply not enough counts in the background regions of any given exposure to get a truly good estimate of the background.

(1) The first item is in fact negligible.
As mentioned above, dark current is by far the dominant source of background, more than five orders of magnitude larger than earthshine and Zodiacal light.
Scattered light from strong airglow lines has no measurable impact, since very black Gunn-Peterson troughs are as black during the day as at night.

(2) While it is true that the background in airglow regions as reported by CALCOS includes a few airglow photons as well as dark current, we cannot use regions contaminated by airglow anyway, and thus this effect is unimportant.
The background is smoothed over 100 pixels in $x$, which will leave some very small contamination beyond the airglow lines, but our cuts around the lines are generous.

(3) When working with very dark spectral regions, $y$-dip is a serious concern.
Due to lowered sensitivity in exposed regions (the primary science aperture), background counts in this region on average have smaller pulse-height amplitudes (PHAs) than those in the detector region used to estimate the background.
Counts in science data are cut on PHA, which greatly reduces the background noise from detector dark current (at low PHA) and cosmic rays (at high PHA).
This means that estimates of the background may in fact be too high by up to $\sim$10\%.
This has little to no effect on most work done with COS, but is important when dealing with dark regions in spectra, such as our \ion{He}{2} Gunn-Peterson troughs.
We discuss this issue in detail in Appendix~\ref{app:bg}, and the effect is shown in Figure~\ref{fig:dark_y} in the Appendix.

(4) Assuming a constant background across all wavelengths is incorrect, because of background variation in the dispersion ($x$) direction.
COS consists of two detectors, A and B segments, and observations in G140L 1105 central wavelength use only the A segment.
(G140L 1280 central wavelength uses both segments, with B segment covering $\lambda < 1150$--1210~\AA, depending on the FP-POS setting, and A segment recording longer wavelengths.)
The background variation in $x$ is fairly smooth on the A segment, which is the only segment used for the majority of our data.
Only SDSS0915 has data from both G140L central wavelengths, and thus from B segment.
In addition, CALCOS takes this variation into account, because these features in $x$ persist over a wide range of $y$, including both the science extraction region and the background estimation region.
See Appendix~\ref{app:bg} for further discussion.

(5) Assuming a constant background also produces compromised results because of the time variability of the background.
This effect is taken into account in the current CALCOS method of processing the background, where each individual exposure is used to estimate its own background.
See Appendix~\ref{app:bg} for further discussion.

(6) Along with $y$-dip, poor statistics due to the low number of background counts is the major drawback of the current CALCOS method of background estimation.
We discuss how to avoid this problem in Appendix~\ref{app:bg}.

We can directly verify background subtraction by consulting other COS quasar data (not taken for \ion{He}{2} studies) where we know that the true flux must be zero for a portion of the spectrum.
The best way to do this is to consider spectra with known low-redshift DLAs or sub-DLAs, where the column density can be constrained by Lyman-series lines to be high enough that the Lyman limit should be black.
Several such examples have been studied with COS/G130M \citep[e.g.,][]{meiring11}, but for them to be of use, we require that some of the Lyman trough be redshifted into the G130M band; i.e., $z_{\rm LLS}>0.24$.

From {\it HST} GO 11598 and 12248 (PI Tumlinson), we have two medium-resolution observations of systems with $N_{\rm H\,I}>10^{19}$~cm$^{-2}$ (implying $\tau_{\rm LL} > 60$, with the column measured from damping wings of line absorption) that are also at sufficiently high redshift to show trough absorption on the B segment in G130M mode.
We calculate the optical depths using the method described in \citet{syphers11b}; we briefly review it here.
This method tallies the number of source counts and background counts in the specified region of the absorption trough for all exposures, and calculates a Poissonian Feldman-Cousins confidence interval \citep{feldman98} for the source counts.
It then takes the power-law continuum fit for the quasar, and calculates the expected number of counts in the specified region absent any absorption, taking into account various detector features like shadows from the grid wires \citep{osterman11}.

To provide some idea of how sensitive a measurement we can accurately perform, we examine the sensitivity as defined by \citet{feldman98}, which is the average upper limit of flux confidence intervals with the given detector background and zero signal.
We also quote the detector upper limit, which is the maximum intensity a source can have without having a probability of $\beta$ of being detected at a significance level $\alpha$ \citep{kashyap10,syphers11b}.
Here we use $\alpha=0.32$ (68\% confidence) and $\alpha=0.05$ (95\% confidence), and $\beta=0.9$.
These flux upper limits become lower limits in optical depth.

The first absorption trough gives a Lyman-limit opacity of $\tau_{\rm LL}=8.8^{+1.6,\infty}_{-0.8,1.2}$, where the errors quoted are 68\% and 95\% confidence, respectively.
The sensitivity is $\tau=8.5$ (68\%) and $\tau=7.9$ (95\%), and the detector lower limits are $\tau=8.1$ (68\%) and $\tau=7.5$ (95\%).
The second absorption trough is oversubtracted, with an estimated $-194$ source counts and $1431$ background counts ($\simeq$65,100 expected source counts if unabsorbed).
This implies $\tau_{\rm LL} \gg 10$, but this is far enough into the nonphysical regime of negative signal that the confidence intervals are highly untrustworthy, because the assumption that the background is correctly estimated is likely invalid.
In this case, we have sensitivities $\tau=8.0$ (68\%), $\tau=7.3$ (95\%), and detector lower limits $\tau=7.6$ (68\%) and $\tau=7.1$ (95\%).

In addition to these two troughs, {\it HST} GO 12204 (PI Thom) provides us with a highly saturated DLA Ly$\alpha$ absorption line ($N_{\rm H\,I}>10^{20.2}$~cm$^{-2}$).
Although the region of zero flux in a line is much shorter than a Lyman-limit trough would be, our using the line allows us to consider this lower-redshift ($z \sim 0.2$) absorber where we do not have coverage down to the Lyman limit.
Fortuitously, the line falls on the A segment in G130M mode and the B segment in G160M mode, giving better statistics than expected, and allowing us to assess both detector segments.
We can find the column density of the absorber quite well from the damping wings of Ly$\alpha$, and convolving this Voigt profile with the COS LSF allows us to limit our analysis to those regions where the flux should truly be indistinguishable from zero ($\tau_{\rm eff}>20$, a residual flux $f_{\lambda} < 10^{-22}$~erg~s$^{-1}$~cm$^{-2}$~\AA$^{-1}$, 5--6 orders of magnitude smaller than dark current).
When we do so, we find the background is overestimated by $\simeq$8\%, yielding a slightly unphysical estimate of source counts, but not extremely so.
The estimates of the optical depth are $\tau \in [8.91,\infty)$ (68\%) and $\tau \in [7.55,\infty)$ (95\%).
Given the background levels, the sensitivities are $\tau=7.5$ (68\%) and $\tau=6.9$ (95\%), while the detector lower limits are $\tau=7.1$ (68\%) and $\tau=6.6$ (95\%).
The data for A segment (G130M) and B segment (G160M) independently agree.
While both are slightly oversubtracted, the difference between the estimated observed counts and the true value of zero is not very significant---if we assume a background at this level estimated with no bias, an oversubtraction at the observed level is expected $\simeq$20\% of the time for the combined segments.

With only three systems satisfying our requirements for observing known-black troughs, we cannot reach firm conclusions.
We do find a hint that oversubtraction of the background is an issue, which is to be expected given the $y$-dip problem.
Alternate methods of determining the background, discussed in Appendix~\ref{app:bg}, are not yet mature enough to use in practice, but promise to yield noticeably more accurate background estimates.

\section{\ion{He}{2} Quasar Results}
\label{sec:results}

We deredden the spectra with a Fitzpatrick \& Massa UV extinction curve \citep{fitzpatrick99}, using $E(B-V)$ from \citet{schlegel98}, and then fit each quasar continuum using a power law.
The fit is iterative, clipping data with a residual above and below threshold values, and the clipping of low data is more aggressive in an attempt to avoid absorption.
For the first iteration we manually specify portions of the spectrum to use, avoiding obvious absorption and emission lines (the latter primarily geocoronal, although some may be intrinsic to the quasars).
When fitting quasars with low-redshift partial LLS, where we can see the break in our COS spectrum but there is substantial flux below it, we fit the spectral energy distribution (SED) including an LLS.
These fits are convolved with the COS LSF, and overplotted in Figures \ref{fig:spec1}--\ref{fig:spec4}.
Higher-redshift LLS, which have breaks in the NUV and are thus typically not detected, are not a substantial concern when extrapolating continua for Gunn-Peterson optical depth measurements.
While it is true that a powerlaw continuum does not exactly remain a powerlaw below a Lyman limit, the deviations are well within our uncertainties.
Even for a high-S/N COS observation, such as that of HE2347, a powerlaw remains a good fit in the FUV despite the intervening high-redshift LLS \citep{shull10}.

We use night-only data in regions contaminated by geocoronal line emission, where possible.
Because most of the observations were a single orbit, sometimes there is no night data available for a given quasar.

With data of this S/N, the optimal approach towards continuum fitting is unclear.
We avoid strong absorption systems, but we have no hope of avoiding the Ly$\alpha$ forest altogether, given our resolution of $\sim$100--150~km~s$^{-1}$.
With $\sim$7~pixels per resolution element, the COS data are also very oversampled.
We bin our higher-S/N data to 7~pixels, with the remaining data binned to 14~pixels, or in one case 28~pixels.
Even for high-S/N data there is rarely reason to bin more finely than 7~pixels, as optimal binning for detecting weak unresolved features is no smaller than this (Keeney et al.\ 2012, in prep.).

We derive uncertainties on our continuum fits by varying the binning of the data, the wavelength range considered, and the clipping level for both high and low excursions.
\citet{worseck11a} obtained continuum uncertainties by assuming a Poissonian distribution of counts, and using a Monte Carlo to sample from this distribution and refitting the spectrum many times.
However, we find that the fitting input parameters such as the data bin size contribute more significantly to the uncertainty than does the spectrum noise.
From each set of these parameters we obtain a normalization and a power-law index; our continuum uncertainty at a specific wavelength is taken to be the full range of model predictions at that wavelength.
Even considering the iterative nature of the fit, there is unfortunately a substantial level of subjectivity in what constitutes ``reasonable'' binning and clipping levels.
For higher-S/N data we typically consider binning the data by 6--14 pixels, 7 being our preferred value for the best estimate, while for lower-S/N data we tend to use $\sim$7--20 pixels, with 14 preferred.
Wavelengths included in the continuum fit typically extend from the flux break (either \ion{He}{2} Gunn-Peterson or low-redshift \ion{H}{1} Lyman limit) redward until we reach wavelengths where the average S/N per resolution element is approximately 1.
For some quasars this value is reached around 1900~\AA, while for others it is closer to 1700~\AA.
We clip fairly aggressively, using $2.5$--3$\sigma$ to clip upward excursions and $1.5$--2$\sigma$ for downward excursions, with $\sigma$ being the local spectrum error.

The observations are detailed in Table~\ref{tab:obs}, the columns of which are as follows:
Column~1---quasar name.
Column~2---right ascension in J2000, from SDSS (even for those quasars discovered in other surveys).
Column~3---declination in J2000, from SDSS.
Column~4---FUV flux at the break wavelength, whether the break is from \ion{He}{2} Gunn-Peterson absorption or a low-redshift \ion{H}{1} LLS.
This is calculated from our spectrum fits, so it applies to the dereddened spectra.
Our Galactic extinctions are typically $E(B-V) \sim 0.01$--$0.03$, which means the flux reported in this table is $\sim$10--30\% higher than the observed flux.
Column~5---the powerlaw index from our fit, taking $f_{\nu} \propto \nu^{\alpha_{\nu}}$. Note $\alpha_{\lambda}=-2-\alpha_{\nu}$.
For those objects with observed LLS where we have usable data above the break, we report $\alpha_{\nu}$ for the SED prior to the onset of the LLS.
Considering the large gap in wavelength coverage between our FUV spectrum (which extends to $\sim$1800--1900~\AA) and the SDSS optical spectrum (beginning at $\sim$3800~\AA), it is likely that several of our quasars have intervening LLS that are not accounted for, and thus this number cannot be directly interpreted as the intrinsic extreme-UV spectral index.
Column~6---exposure time.
Column~7---observation date.
Column~8---the optical depth in a 20~\AA\ bin (observed frame) just shortward of the \ion{He}{2}~Ly$\alpha$ break, or as near to that as possible while avoiding geocoronal emission and any obvious line-of-sight proximity effect. The uncertainties reflect both the number of observed counts in the trough (68\% limits) and the full range of SED fits plotted in Figures~\ref{fig:spec1}--\ref{fig:spec4}.

As shown in Table~\ref{tab:obs}, average effective \ion{He}{2} Ly$\alpha$ optical depth does evolve from $\tau_{\rm eff} \sim 2$ at $2.7 \lesssim z \lesssim 2.9$ to $\tau_{\rm eff} \gtrsim 5$ at $3.2 \lesssim z \lesssim 3.6$, as expected if we are probing the end of helium reionization at $z \sim 3$.
There is considerable variation in optical depth between sightlines and even in a single sightline, but when averaging over larger bins to find effective optical depths, these fluctuations are reduced, and we see steady progression to lower optical depth at lower redshift.
Current model predictions for the evolution of $\tau_{\rm eff}$ with redshift differ considerably \citep{mcquinn09,dixon09,worseck11a}, and it will be interesting to see how more accurate simulations with radiative hydrodynamics affect these results \citep{meiksin11}.
Other aspects, such as quasar beaming and SEDs, can also noticeably affect results.

Comments on the spectra:\\
{\it 4C57.27}---the high flux of this target allows us to use data out to 1950~\AA, and verify an LLS at $z=0.9615$ (with the redshift from Ly$\delta$ and Ly$\epsilon$), with $\log{N_{\rm H \, I}} \simeq 17.3$. Due to the fairly poor quality of data above the Lyman limit, we fit a power law to the data below the break.\\
{\it SDSSJ1443+3546}---the LLS at $z=0.444$ has $\log{N_{\rm H \, I}}=17.24 \pm 0.02$.\\
{\it SDSSJ0947+1421}---there is a substantial flux dip from $\sim$1370--1410~\AA, but the origin is unclear. It appears in both COS exposures at the same position, despite the FP-POS shift between exposures of $\sim$20~\AA, and does not appear in spectra of other targets. It therefore seems unlikely to be an instrumental feature. Assuming the dip is \ion{H}{1}~Ly$\alpha$ absorption, the rest-frame equivalent width of the feature is $15.3 \pm 0.3$~\AA, which would make it a DLA, but it would have to have partial covering or contaminating flux from a nearby source to explain the fact that it is not saturated at line center. Possible dips near 1500~\AA\ and perhaps even 1600~\AA\ suggest something unusual may be evident in this spectrum, or something amiss with this observation.\\
{\it SDSSJ1244+6201}---the LLS at $z=0.528$ has $\log{N_{\rm H \, I}}=17.25 \pm 0.10$. The large flux rise at $\lambda \sim 1130$~\AA\ is geocoronal \ion{N}{1}~$\lambda$1134. Other relatively unusual geocoronal lines clearly visible are \ion{O}{1}~$\lambda$1356 and \ion{N}{1}~$\lambda$1200 (visible in the error spectrum). No night data is available for this target.\\
{\it SDSSJ0856+1234}---the LLS at $z=0.77$ has $\log{N_{\rm H \, I}}=17.23 \pm 0.04$.\\
{\it SDSSJ1508+1654}---this target has a very uncertain redshift. SDSS reports $z=3.17$, but this is skewed by high-ionization lines. \ion{O}{1}~$\lambda$1302 emission gives a redshift of $z \simeq 3.19$, while \ion{C}{4} is $\sim$3000~km~s$^{-1}$ blueshifted from this. \ion{H}{1}~Ly$\alpha$ absorption starts at $z=3.21$, which matches the onset of \ion{He}{2}~Ly$\alpha$ absorption seen in the COS spectrum, and is therefore adopted here. If the hydrogen and helium absorption is due to an infalling system near the quasar, it is one that shows no \ion{N}{5} or \ion{C}{4} absorption in the SDSS spectrum ($R \simeq 1740$, S/N$\,\simeq 30$ per resolution element).\\
{\it SDSSJ1025+0452}---the LLS at $z=0.52$ has $\log{N_{\rm H \, I}}=17.93^{+0.12}_{-0.07}$ (68\% confidence interval; note that an infinite column density is in the 95\% confidence interval). There is possible \ion{Mg}{2} absorption near this redshift in the SDSS spectrum, but \ion{H}{1}~Ly$\alpha$ absorption precludes robust identification at SDSS resolution.\\
{\it SDSSJ0855+2932}---the LLS at $z=0.555$ has $\log{N_{\rm H \, I}}=17.74^{+0.11}_{-0.07}$ (68\% confidence interval; $<18.12$ at 95\% confidence). There is a plausible strong \ion{Mg}{2} absorber at $z=0.5565 \pm 0.0001$ in the SDSS spectrum, although it is blended with the \ion{H}{1}~Ly$\alpha$ forest.\\

One quasar, SDSSJ0054+0028, was found to not be FUV bright---although flux $\sim$$5 \times 10^{-18}$~erg~s$^{-1}$~cm$^{-2}$~\AA$^{-1}$ is observed in the COS spectrum, it continues unabated below the \ion{He}{2} break.
Thus the quasar must lie behind the faint, low-redshift galaxy barely visible in the SDSS image, which is the source of the UV light in the {\it GALEX} photometry, as well as the COS spectrum.

\subsection{Match Rate Statistics}
\label{sec:match_rate}

Our current survey found 13 new \ion{He}{2} quasars out of 16 targets (and 15 UV detections of the target quasars).
We thus find that 94\% (85--98\%) are FUV bright, and 81\% (70--89\%) are \ion{He}{2} quasars.
(We quote 68\% Wilson confidence intervals for these quantities; \citealt{brown01}.)
Our earlier reconnaissance programs with the ACS prism yielded \ion{He}{2} quasars at a rate of 58\% (50--66\%, 22 out of 38 observations) for targets in our SDSS--{\it GALEX} cross-correlation catalog \citep{syphers09b,syphers09a}.
Objects in this catalog are known quasars with flux detected in one or both {\it GALEX} bands, and an SDSS--{\it GALEX} separation $r<3\arcsec$, as discussed in the previous section.
We estimated that a success rate of $\sim$80\% could be achieved if one required detection in both {\it GALEX} bands, and this is borne out by the current study.
Requiring flux in one band only occasionally led to no FUV flux observed at all, either due to the source being spurious or having only NUV flux (due to a higher redshift LLS).
Our current survey finds true UV flux in all cases, although in one instance this was not due to the quasar, but rather a low-redshift galaxy very near the quasar sightline.

Cases such as the interloping low-redshift UV source, or those quasars with low-redshift LLS, show why short reconnaissance spectra are still advisable, despite the vast increase in efficiency that the SDSS--{\it GALEX} cross-correlation technique has brought.

\subsection{Quasar Emission}

\begin{figure}
\epsscale{1.0}
\plotone{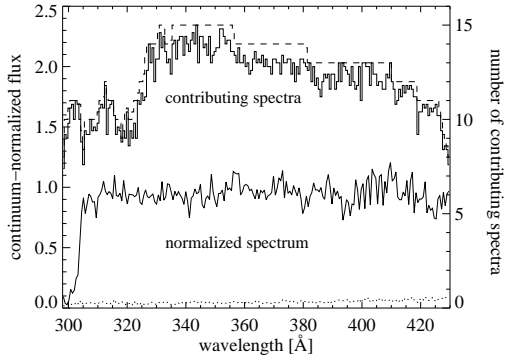}
\caption{Rest-frame stack of all \ion{He}{2} quasars presented in this paper, binned to $0.6$~\AA. The lower two curves are the coadded spectrum (solid line) and the error on this spectrum (dotted line), while the upper two curves are the total number of spectra allowed to contribute at a given wavelength (dashed line) and the number of spectra actually contributing at a given wavelength after clipping (solid line). The spectra are all normalized by their power-law fits (as shown in Figures~\ref{fig:spec1}--\ref{fig:spec4}) before coaddition, so no information about spectral index remains, but emission lines should be evident if present. None are seen at high significance, despite theoretical predictions of a \ion{He}{2}~Ly$\alpha$ EW$\sim$10~\AA. The flux break at 304~\AA\ is due to the onset of IGM \ion{He}{2} absorption.}
\label{fig:stack}
\end{figure}

To see if there is any evidence of extreme-UV (EUV) quasar emission lines in our data, we normalize all the spectra by their continuum fits as shown in Figures~\ref{fig:spec1}--\ref{fig:spec4}, and stack them in the rest frame using a robust mean (clipped at $3\sigma$).
The resulting average spectrum is shown in Figure~\ref{fig:stack}, where the error spectrum (dotted line) is calculated directly from the standard deviation of all contributing spectra.
Also plotted in this figure is the number of spectra that can contribute at each wavelength (the dashed upper curve), which depends on excisions of data around geocoronal emission and truncation due to observed-frame wavelength coverage.
Because we clip the data at every wavelength to avoid extreme noise excursions and absorption features, some additional spectra are excluded, and so the total number actually used for the average at each wavelength is plotted as well (the solid upper curve).
The normalization means no information about the EUV spectral index remains, but the stack would reveal common strong emission lines if they were present.
In fact we see only marginal hints of lines, and notably we see no evidence at all for \ion{He}{2}~Ly$\alpha$ emission.

Simple Cloudy \citep{ferland98} photoionization models of the quasar broad emission lines region (BELR) predict very noticeable \ion{He}{2}~Ly$\alpha$ emission, with an equivalent width of 6--11~\AA\ for a covering factor of 10\% \citep{syphers11a}.
Very low-resolution ACS prism spectra occasionally showed lines that might be \ion{He}{2}~Ly$\alpha$ \citep{syphers09b,syphers09a}, but this was far from universal.
It is notable, therefore, that {\it none} of the COS spectra presented in this paper show any obvious emission lines, and no hint of \ion{He}{2}~Ly$\alpha$ emission is seen even in stacks of the data.
\citet{lawrence11} has recently proposed an extreme-UV reprocessing region in AGN to create the Big Blue Bump, which produces \ion{He}{2}~Ly$\alpha$ even more strongly than standard BELR models.
The observed absence of the line may present a strong constraint on this reprocessing, in addition to presenting a challenge for understanding even the standard broad line region.

The quasar EUV spectral index is also of great interest, as it is these photons that create and destroy many important ions in the IGM (\ion{He}{2}, \ion{Si}{4}, \ion{C}{4}, \ion{O}{6}, etc.).
Constraining quasar emission in this ionizing regime is important for determining the ionizing UV background \citep{haardt11} and models of helium reionization \citep{meiksin05}.
\citet{telfer02} find an average index $\alpha_{\nu}=-1.76 \pm 0.12$, while \citet{scott04} find $\alpha_{\nu}=-0.56^{+0.38}_{-0.28}$ using a lower-redshift, lower luminosity sample.
However, both these samples use data only in the energy range $1 < E \lesssim 1.5$~Ryd, whereas our FUV spectra of $z \sim 3$ quasars allows us to examine $1.8 \lesssim E < 3$~Ryd.
We caution that the power-law indices quoted in Table~\ref{tab:obs} are affected by the Lyman Valley \citep{moller90} and $z \gtrsim 1$ LLS that have their Lyman limits in the NUV, and are thus unobserved.
Both these effects bias these indices to be higher (harder) than the actual intrinsic spectrum.
It is nonetheless interesting that none of our quasars has a best-fit spectral index as soft as the \citet{telfer02} average, and only one is consistent with that average.
It is true, however, that some of the \ion{He}{2} quasars observed with ACS had redder slopes, although extremely low resolution made it difficult to define the continuum \citep{syphers09b}.
As the sample of \ion{He}{2} quasars increases, it will be interesting to apply statistical corrections for IGM opacity and get a direct measurement of the average EUV spectral index.

\section{Conclusion}
\label{sec:conclusion}

Quasars allowing the study of \ion{He}{2} reionization through examination of their Gunn-Peterson troughs are still relatively rare---there are 33 in the literature \citep[e.g.,][]{zheng05,syphers09b,syphers09a,worseck11a}.
Many of these are fairly faint, $f_{\lambda} \sim 10^{-17}$~erg~s$^{-1}$~cm$^{-2}$~\AA$^{-1}$.
In this paper we present 13 new \ion{He}{2} quasars, including several that are quite bright compared to those previously cataloged ($f_{\lambda} > 2 \times 10^{-16}$~erg~s$^{-1}$~cm$^{-2}$~\AA$^{-1}$; i.e., brighter than the well-studied Q0302$-$003), and well suited for higher-resolution, higher-S/N COS spectroscopy.
These initial observations show evidence for IGM helium opacity increasing with redshift, as expected during the end of the helium reionization epoch.
Higher-resolution observations will allow detailed characterization of the transition from Gunn-Peterson troughs to the \ion{He}{2}~Ly$\alpha$ forest, and allow us to move beyond effective optical depth methods to methods such as dark gap and transmission window statistics \citep[e.g.,][]{gallerani08,maselli09}.

We see no clear EUV quasar emission lines either in individual quasars or in an averaged spectrum.
Because models of EUV quasar emission generally predict easily observable \ion{He}{2}~Ly$\alpha$ emission, it is interesting that we see no sign of it.
This could provide useful constraints on models of the broad line region in AGN, as could any EUV metal line emission seen in future studies of individual higher-quality spectra or averages of larger samples.

COS has allowed a revolution in the observation of the end of the \ion{He}{2} reionization epoch.
The new COS G130M/1220 mode allows study of the helium IGM at higher resolution ($R \gtrsim \;$10,000) down to $z=2.5$, and along with G140L, it will allow observation of Gunn-Peterson troughs of higher-order Lyman lines \citep{syphers11b}.
COS data should nonetheless be used carefully, as there are a number of detector features that become important when dealing with very faint targets or Gunn-Peterson troughs.
These issues may not be resolvable for the average COS user, and are not dealt with in CALCOS 2.13.6 or 2.15.4 (or development CALCOS versions up to 2.16; S.\ Penton 2011, private communication), but they are manageable problems that future versions of the CALCOS pipeline should handle.

Cross-matching optical quasar catalogs with {\it GALEX} UV catalogs has previously been shown to be a very efficient method for finding \ion{He}{2} quasars \citep{syphers09b,syphers09a}.
In this paper we have robustly demonstrated that adding the requirement that the quasar be detected in both {\it GALEX} bands raises the efficiency to $\sim$80\%.
Large expansions of quasar catalogs, in particular SDSS DR7 \citep{schneider10}, have allowed us to use this more selective criterion, and future catalogs resulting from, e.g., Pan-STARRS \citep{kaiser02} and BOSS \citep{eisenstein11}, will allow us to require not only matching in both bands, but also a minimum FUV flux.
\ion{He}{2} quasar studies are starting to allow detailed characterization of the progress of helium reionization, and possible future observations promise a great deal more progress.

\acknowledgments

We thank Jason Tumlinson and Chris Thom for providing access to COS observations of high-column, low-redshift absorbers from {\it HST} GO 11598 and 12248 (PI Tumlinson) and GO 12204 (PI Thom).
We thank Steve Penton and Steve Osterman for helpful discussions about the COS detector.

Support for {\it HST} Program number 12178 was provided by NASA through grants from the Space Telescope Science Institute, which is operated by the Association of Universities for Research in Astronomy, Incorporated, under NASA contract NAS5-26555.

The Institute for Gravitation and the Cosmos is supported by the Eberly College of Science and the Office of the Senior Vice President for Research at the Pennsylvania State University.


\appendix

\section{The COS FUV Background}
\label{app:bg}

We have used 532~ks of dark data taken from October 2009 through October 2011 to characterize the COS FUV cross delay line (XDL) microchannel plate (MCP) detector background.
These data are from the calibration programs {\it HST} GO 11895 and 12423.
We use CALCOS 2.13.6 for our data in this paper, and this version has been in widespread use since December 2010.
As of this writing, CALCOS 2.15.4 has recently been released (although currently it is neither used by the STScI archive nor widely used elsewhere), so we also discuss that version, and include some information on development versions of CALCOS.

The primary science aperture (PSA) spectrum is roughly located at $y \sim 462$--482 on segment A and $y \sim 510$--530 on segment B, where $y$ is the cross-dispersion direction.
(Note that the FUV detector does not have physical pixels---digitized positions are derived for each count, and it is these that we consider as ``pixels.''
These raw coordinates translate to geometrically corrected coordinates of roughly $y \sim 485$--505 on segment A and $y \sim 545$--565 on segment B.)
The mean background in those PSA regions can be calculated quite precisely, and gives an estimate of what to expect on average: $(1.959 \pm 0.004) \times 10^{-6}$~counts~s$^{-1}$~pixel$^{-1}$ on segment A, and $(1.932 \pm 0.004) \times 10^{-6}$~counts~s$^{-1}$~pixel$^{-1}$ on segment B, where here we refer to two-dimensional pixels.
However, there are important spatial and temporal variations in the background rate, which cannot be ignored.

\begin{figure}
\epsscale{1.0}
\plotone{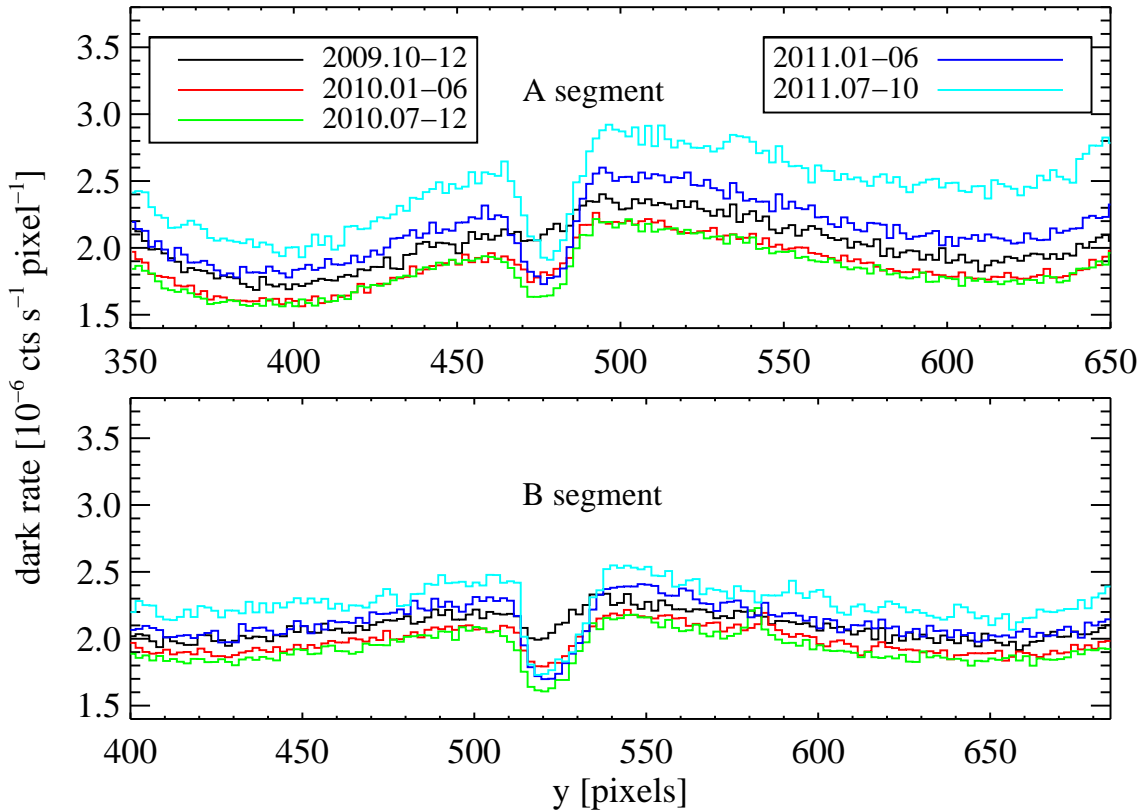}
\caption{Dark count rates for two-dimensional pixels on the COS detector, as a function of $y$ (the cross-dispersion direction), binned by two $y$ pixels. We have marginalized over $x$ for $1500 < x < \;$15,000. The $y$-dip is evident on both segments, occurring where the spectrum is located ($y \sim 475$ for A segment, and $y \sim 520$ for B segment). Since the sensitivity varies with time, we plot the dark data binned by 3--6 months. We use pulse-height filtering, selecting those counts with $2 \le {\rm PHA} \le 30$. Note that the dark counts are averaged using raw coordinates; the dip is still evident, though less clear, in corrected coordinates.}
\label{fig:dark_y}
\end{figure}

The temporal variation has three main components.
First is the instrument settings.
The dark rate depends on the high-voltage (HV) setting, which was initially high when COS was installed in May 2009, lowered in August 2009 to reduce background features, and then increased again only for B segment in March 2011 to counteract gain sag.
The lowering of the HV substantially decreased the background; we use no science data from before this change, and therefore do not consider that earlier time period here.
The HV increase, being only for segment B, does not affect G140L/1105 data, and in addition, it had a very small impact on the background.
The second variation is gain sag itself.
The MCP detector becomes less efficient in converting photons to counts the more it is used \citep{dixon10}.
For every count, COS records an associated pulse-height amplitude (PHA), ranging from $0$ to $31$.
Many background or contaminating counts have very low or very high PHAs, and thus the CALCOS pipeline defaults to using only those counts with $2 \le {\rm PHA} \le 30$.
(These limits have varied; the lower limit was previously 4, although the observer has access to all ${\rm PHA} > 0$ counts and can select their own range.)
As an example, the average dark rate of $(1.959 \pm 0.004) \times 10^{-6}$~counts~s$^{-1}$~pixel$^{-1}$ on segment A is for a cut $2\leq \;$PHA$\; \leq 30$, but is reduced to $1.714 \times 10^{-6}$~counts~s$^{-1}$~pixel$^{-1}$ if the cut is $4\leq \;$PHA$\; \leq 29$.
Gain sag causes PHAs to slowly decrease over what they would formerly have been, and when events start falling below the PHA cutoff, they are discarded.
This causes ``$y$-dip'' in the spectrum, shown in Figure~\ref{fig:dark_y}.
The progression of this problem with time is evident.
Note, however, that $y$-dip does not affect all areas of the spectrum equally---the effect is concentrated at those $x$ positions where the bright Ly$\alpha$ airglow line falls in various combinations of central wavelength and FP-POS.
Third, on much shorter time scales, the dark rate depends on the spacecraft environment, which can temporarily increase the background by about a factor of two.
(Data is not taken in the SAA, where the background can be an order of magnitude larger.)

\begin{figure}
\epsscale{1.0}
\plotone{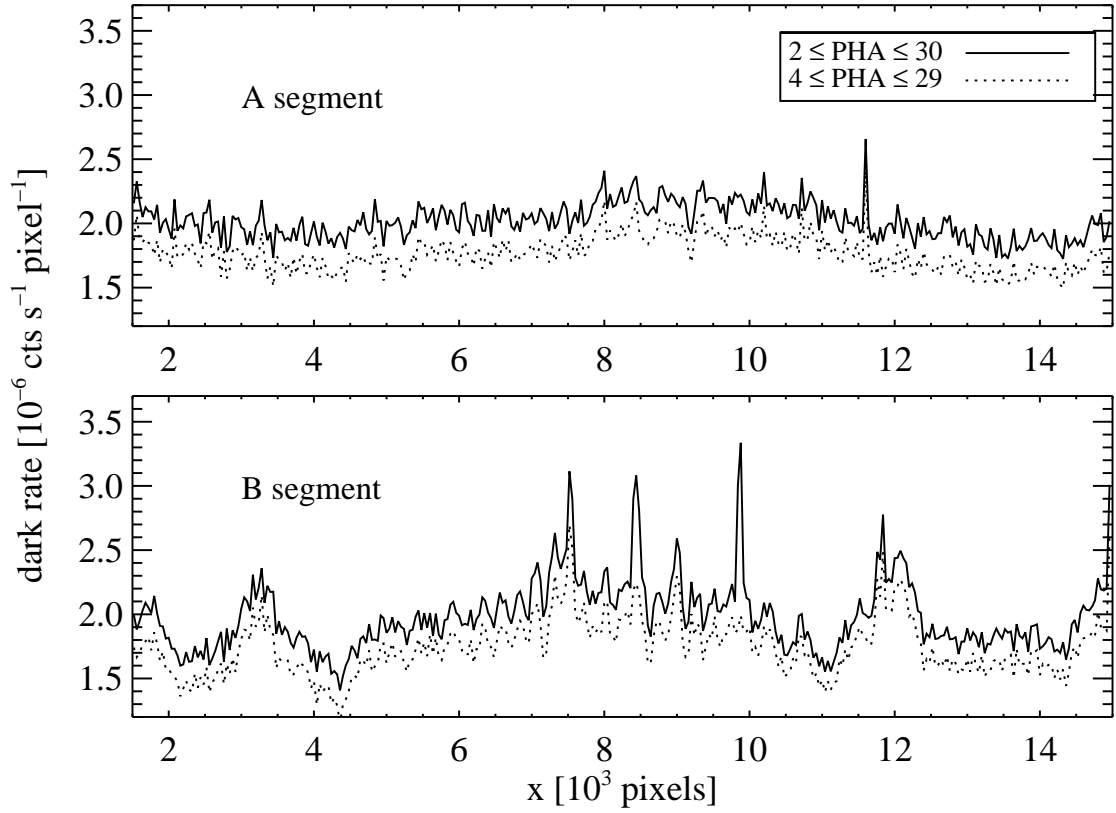}
\caption{Dark count rates for two-dimensional pixels on the COS detector, as a function of $x$ (the dispersion direction), binned by 40 $x$ pixels. We have marginalized over $y$ within $462 < y < 482$ for segment A and $510 < y < 530$ for segment B, which is about where the PSA spectrum falls. Segment A is largely free of narrow features, while segment B shows prominent features, even though we use pulse-height filtering. These features are reduced when we restrict to those counts with $4 \le {\rm PHA} \le 29$, and are substantially worse if no PHA filtering is imposed. These features do not vary appreciably with time. Note that the dark counts are averaged using raw coordinates.}
\label{fig:dark_x}
\end{figure}

There is also spatial variation in the background in the dispersion direction, shown in Figure~\ref{fig:dark_x}.
The variation on A segment is smooth, although at $\sim$20\% it is not inconsequential.
B segment clearly shows features, and while PHA filtering removes the strongest features (an order of magnitude stronger than those shown here), the remaining features are non-negligible.
In an attempt to mitigate the loss of sensitivity of the detector, the default lower limit on PHAs has been reduced from 4 to 2, but this has made some of the features on B segment much stronger (those counts with PHA=30 are not significant).
Note that CALCOS 2.13.6 incorrectly flags $x$ positions as being in wire shadows even in the background windows, which are not exposed and thus unaffected by the wires.
This leads to spurious dips in the background spectrum as reported by CALCOS, though this is fixed in versions 2.16 and later (S.\ Penton 2011, private communication).

The method of dealing with the background used by CALCOS versions 2.16 and earlier is to calculate an average of the background in two regions symmetrically offset in $y$ from the science spectrum and incorporating a total of 96 $y$ pixels, and smooth that in the $x$ direction by 100 pixels.
Considering the average background rate and exposure time ($\sim$1--3~ks), one problem is evident---there are on average only 18--56 counts in the region used to estimate the background for a one-dimensional pixel, too few to determine the background accurately (excursions $>$20\% from the true background will occur 40\% of the time for 1~ks exposures).
For Gunn-Peterson troughs, where we are averaging over hundreds of one-dimensional pixels, this is a less important problem (a trough 20~\AA\ in the observed frame will have a background $>$5\% off only $\sim$0.2\% of the time, but $>$2\% off $\sim$20\% of the time).
More worrisome is that the background is determined from regions not subject to $y$-dip, unlike the spectrum extraction region.
A much better method is to create a detailed two-dimensional map of the dark current using all available dark data for a certain time period, scale this to the overall dark rate determined in an exposure (which can vary from the mean due to the local environment of {\it HST}), and use the result for calculating the background.
This approach is currently being implemented in a new version of CALCOS (S.\ Penton 2011, private communication).
Lifetime position shifts, where the PSA is moved to a previously unexposed portion of the detector, are part of the design of COS.
Such a shift will move the PSA out of the current $y$-dip, but a new dip will be created, and the improved background technique discussed here will still be needed.
Frequent monitoring of the background is needed whether or not such shifts take place.

The default primary science aperture extraction window in CALCOS 2.15.4 and earlier is 57 pixels high in $y$ for G140L, and thus a one-dimensional pixel is 57 two-dimensional pixels.
For G130M and G160M, the default is 35 pixels in $y$.
(Note that the COS data handbook and other COS documentation consistently imply that a one-dimensional pixel is 10 two-dimensional pixels; \citealp{dixon10}.
It is unclear why, as this has never been the default for data {\it HST} users receive.
The Exposure Time Calculator\footnote{http://etc.stsci.edu/etc/input/cos/spectroscopic/} specifies the aperture for which it reports the dark current, and this is currently 47~pixels for both medium and low-resolution gratings.)
This default extraction window is a poor choice for those interested in low-flux regions, since it is much larger than the extent of the science spectra.
The PSA has $>$90\% of the source counts within $\simeq$10--20 $y$~pixels, and $>$99\% within $\simeq$20--30 $y$~pixels (as can be roughly seen by the $y$-dip region in figure \ref{fig:dark_y}).
However, this extraction height is adjustable in CALCOS, and running a custom extraction with a more appropriate window allows one to reduce the background by a factor of 2--4 without losing an appreciable number of science counts.

For spectra with regions of very low flux, we therefore recommend this custom extraction with a smaller extraction height, although this must be accompanied by custom flux calibration if the extraction window is small enough to exclude a noticeable number of source counts.
The flux calibration is nontrivial for small extraction heights, because the $y$~extent of the spectrum varies as a function of $x$, and thus the sensitivity curve changes in a wavelength-dependent manner, rather than a simple multiplicative constant.
Ideally one would also trace the spectrum across the detector, as it does not fall exactly on a constant $y$ all the way across, but the variation is small, and such tracing is not possible with current versions of CALCOS.
We caution that a narrower extraction window could exacerbate the $y$-dip problem with the current background method, since the PSA will no longer include larger regions outside of the dip.

As an example of the advantage of a narrower window, we consider an exposure where the default G140L window predicts 500 background counts and $2 \times 10^4$ unabsorbed source counts.
The Feldman-Cousins sensitivity is $26.37$ counts or $\tau=6.63$ at 68\%, and $47.57$ counts or $\tau=6.04$ at 95\%.
If we use an extraction height of 20 pixels, the background is reduced by a factor of $57/20=2.85$ to $175.4$, while the source flux is reduced by only $\simeq 3.5$\% to $1.93 \times 10^4$.
In this case, the Feldman-Cousins sensitivities become $\tau=7.10$ (68\%) and $\tau=6.50$ (95\%).

\end{document}